\documentclass{article}
\usepackage{spconf,amsmath,graphicx}
\usepackage{booktabs, subfig}
\usepackage{physics}

\def\L{{\cal L}}

\title{FINE-GRAINED STYLE CONTROL IN TRANSFORMER-BASED TEXT-TO-SPEECH SYNTHESIS}
\name{Li-Wei Chen, Alexander Rudnicky}
\address{Language Technologies Institute, Carnegie Mellon University\\
\small\texttt{\{liweiche, air\}@cs.cmu.edu}}
\begin{document}
%
\maketitle
\begin{abstract}
\vspace{-0.25em}
In this paper, we present a novel architecture to realize fine-grained style control on the transformer-based text-to-speech synthesis (TransformerTTS).
Specifically, we model the speaking style by extracting a time sequence of local style tokens (LST) from the reference speech.
The existing content encoder in TransformerTTS is then replaced by our designed cross-attention blocks for fusion and alignment between content and style.
As the fusion is performed along with the skip connection, our cross-attention block provides a good inductive bias to gradually infuse the phoneme representation with a given style.
Additionally, we prevent the style embedding from encoding linguistic content by randomly truncating LST during training and using wav2vec 2.0 features.
Experiments show that with fine-grained style control, our system performs better in terms of naturalness, intelligibility, and style transferability.
Our code and samples are publicly available.~\footnote{https://github.com/b04901014/FG-transformer-TTS}

\end{abstract}
\vspace{-0.1em}
\vspace{-1.0em}
\section{Introduction}
\vspace{-0.8em}
\label{sec:intro}
Human speech naturally consists of linguistic content and para-linguistic information.
Para-linguistic information can be further categorized into speaker characteristics and speaking styles.
Speaker characteristics correspond to the vocal properties of a specific speaker, such as accent and voice types.
On the other hand, time-varying properties within a sentence such as speaking rate, loudness and prosody belong to speaking styles.

Separation of speaker characteristics is widely used in neural-based TTS systems~\cite{tacotron,fastspeech,tacotron2,Li2019NeuralSS} to improve controllability and expressivity.
In the global style token (GST)~\cite{Wang2018StyleTU}, a reference encoder is trained to capture speaker characteristics by representing the reference speech with a linear combination of style tokens.
Conditioned on GST, TTS systems can synthesize speech with different speaker characteristics.
Nonetheless, modeling of speaker characteristics and speaking styles often suffers from an issue termed ``content-leakage''~\cite{DBLP:conf/icassp/HuSTD20}.
This is a situation where the system erroneously extracts linguistic information from the reference speech, resulting in a content mismatch between the input text and the generated speech during inference.
Content-leakage stems from minimizing a reconstruction loss of the reference speech during training.
Without additional constraint, it relies solely on the design of the network architecture to prevent encoding linguistic information from the reference speech.

To make speaking styles controllable, Skerry-Ryan et al.~\cite{pmlr-v80-skerry-ryan18a} augment a style encoder that extracts a single prosody embedding from the reference audio.
As fixed-length representation fails to model fine-grained styles, Lee et al.~\cite{Lee2019RobustAF} extend the work to extract variable-length prosody embeddings with a soft attention mechanism for the fusion between text and prosody.
In \cite{Klimkov2019}, the attention mechanism is further replaced by forced alignment of reference speech to aggregate prosody embeddings on a phoneme basis.
Later, Tan et al.~\cite{tan21_interspeech} introduce additional objectives using collaborative learning and adversarial training to mitigate content-leakage.

With the success of transformer based models in various domains~\cite{Devlin2019BERTPO,NEURIPS2020_92d1e1eb,li-etal-2020-bert-vision}, TransformerTTS~\cite{Li2019NeuralSS} is developed as an alternative to seq2seq TTS systems, as it provides faster inference speed without noticeable quality degradation.
Several studies have adapted prior works on modeling speaker and style properties to the TransformerTTS framework.
MultiSpeech~\cite{Chen2020MultiSpeechMT} extends TransformerTTS with an additional speaker module, resembling the approach of GST.
UTTS~\cite{paul21_interspeech} further adapts \cite{pmlr-v80-skerry-ryan18a} to TransformerTTS and minimizes an additional Renyi divergence objective to alleviate content-leakage.
However, fine-grained control of prosody, speaking rate, and other subtle speaking styles remains unexplored for TransformerTTS.

In this paper, we incorporate fine-grained style control into TransformerTTS.
We condition the speaking style on reference speech, making our model infer the style factors where the perceptual difference may be hard to quantify manually.
Motivated by \cite{Lee2019RobustAF}, we develop a cross-attention module to fuse linguistic content and style information into a mixed representation.
We show that our designed model architecture and training procedure suffer less from content-leakage and preserves style information without additional regularization objective.
Moreover, by separately controlling fine-grained style factors, our system generates higher quality speech.

\begin{figure*}
    \centering
    \includegraphics[width=\linewidth]{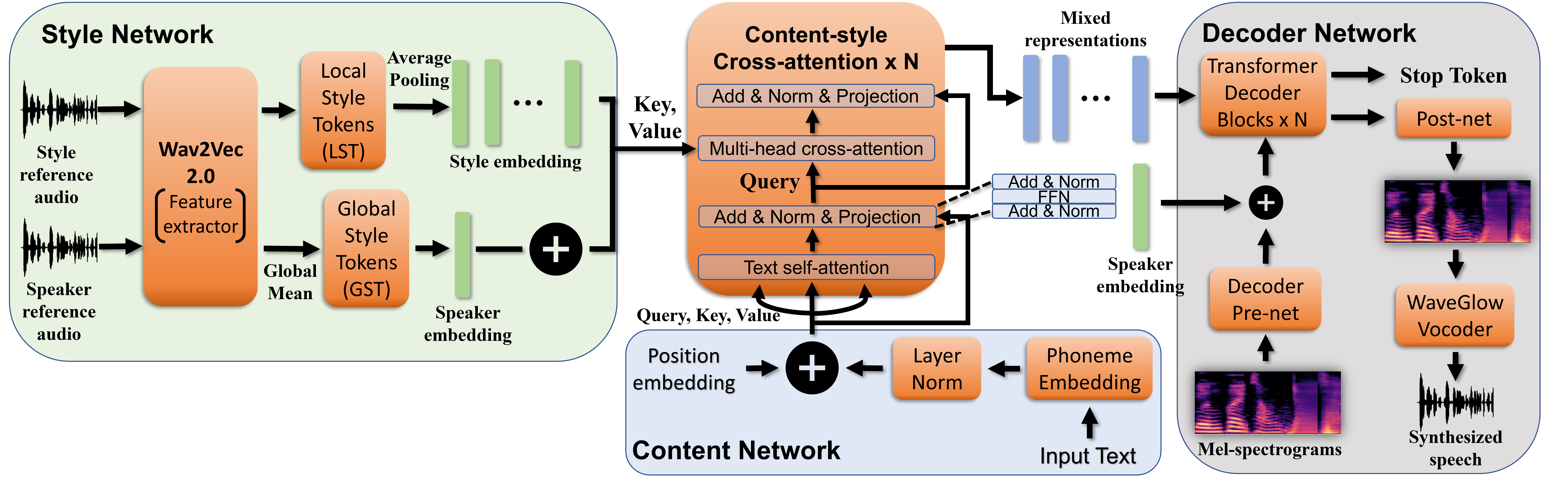}
    \caption{Overview of our LST-TTS system.}
    \label{fig:system}
\end{figure*}

\section{Method}

Figure~\ref{fig:system} shows the overview of our LST-TTS system.
We decompose our system into four sub-modules: content network, style network, decoder network, and the content-style cross-attention blocks.
The decoder network follows an  architecture similar to MultiSpeech~\cite{Chen2020MultiSpeechMT}, while the content network is a simplified version containing only phoneme embedding and layer normalization.
Here we focus on the design of the style network and content-style cross-attention blocks.

\subsection{Style network}
The purpose of the style network is to extract and disentangle speaker characteristics and fine-grained speaking styles from the reference speech.

\textbf{Wav2vec 2.0 features.}
Instead of using spectral based features or pretrained speaker embeddings, we adopt wav2vec 2.0~\cite{NEURIPS2020_92d1e1eb} to extract meaningful representations from the audio.
While wav2vec 2.0 was originally meant to be fine-tuned on downstream tasks, here we use it as a feature extractor.
The benefit this brings is twofold.
The content leakage is mitigated since the reconstruction of reference speech from wav2vec 2.0 features is more difficult compared to raw waveform or spectrum-based inputs.
Moreover, since wav2vec 2.0 is already a transformer based model, its output features are naturally suitable to be processed by the succeeding attention blocks.

\textbf{Speaker embedding.}
For the global style embedding, we adopt a framework similar to GST~\cite{Wang2018StyleTU}.
Specifically, the output features from wav2vec 2.0 are processed by a position-wise linear layer with ReLU activation.
A single layer LSTM~\cite{10.1162/neco.1997.9.8.1735} is applied for further contextualization.
Finally we aggregate a mean representation across time-steps to form a global style query vector.
To obtain the speaker embedding, the query vector will be fed to a multi-head attention~\cite{NIPS2017_3f5ee243}, with a trainable codebook of global style tokens as both key and value.

\textbf{Style embedding.}
Similarly, the output features from wav2vec 2.0 are also processed by a LSTM for fine-grained style embeddings.
Each time step of the output will be fed as a query to a multi-head attention with another trainable codebook as key and value to produce a sequence of preliminary style embeddings.
We term the embedding codebook as ``local style tokens'' (LST), as it is constructed in a similar way with GST but focuses on frame-level properties.
Then instead of taking the mean across the time dimension, we apply average pooling with stride $4$ and kernel size $8$ to smooth out the style embeddings.
The final style representation is obtained by broadcast-adding speaker embedding with the smoothed style embedding across the time dimension.

\subsection{Content-style cross-attention blocks}
In fine-grained style control, one persistent issue is to align variable-length prosody features with the phonetic representations.
In \cite{Lee2019RobustAF}, a scaled dot-product attention is applied to learn the alignment between the two modalities.
The prosody features are then aggregated per phone to concatenate with the phone representation.
However, simple concatenation is a rather naive approach for the fusion of style and content information and is more vulnerable to content-leakage.

With the transformer architecture, we can embed the style information into the linguistic content in a learnable and natural manner.
Specifically, we replace the original content encoder blocks in TransformerTTS with cross-attention blocks.
As shown in Figure~\ref{fig:system}, phone representations are first processed by a typical transformer encoder block.
The output is then used as the query for a multi-head cross-attention module with style representation as both key and value.
The resulting aligned style representation will be added back to the content representation with the residual design in transformer blocks.
Similar to the transformer encoder, this block can be stacked multiple times for better performance.
Since the same style representation is fed as key and value for multiple blocks within the skip connections, one can view the functionality of this module as gradually refining the content representation based on the style representation.


\subsection{Training and inference procedure}
\label{ssec:training}
Our system is trained end-to-end with essentially the same loss as TransformerTTS~\cite{Li2019NeuralSS}.
We denote the style network as $S$, content network as $T$, content-style cross-attention as $A$, and decoder network as $D$.
During training, we random sample a pair of speech and text $(\vb x^s_{sty}, \vb c)$ from speaker $s$, along with an additional utterance $\vb x^s_{spk}$ from the same speaker as the speaker reference audio.
The system is then trained to minimize the following loss:
\[
    \L_{tts} = \| D(A(S(\vb x^s_{sty}, \vb x^s_{spk}), \vb c), \vb x^s_{sty}) - \text{Mel}(\vb x^s_{sty})\|_1
\]
Additionally, a post-net is trained to further refine the output mel-spectrogram, and a stop token is jointly trained to predict the end of the speech.

\textbf{Random truncation of style embedding.} At training time, we truncate the style embedding to a random length $\tilde{l}_{sty}$ chosen from $[\alpha, l_{sty}]$, where $l_{sty}$ is the length of the original embedding and $\alpha$ is a hyper-parameter controlling the minimum length after truncation.
We set $\alpha=15$ in our experiments.
This benefits our system in two aspects.
First, at inference time, $\vb x_{sty}$ may be much shorter than expected synthetic speech as the content is different.
Our designed training procedure simulates this situation.
The model is trained to match the style of synthetic and reference speech up to $\tilde{l}_{sty}$, and fill out the remaining styles based on the given style information before $\tilde{l}_{sty}$.
One can also observe this behavior from the synthesized speech samples we provided.
Furthermore, content-leakage is alleviated as only a partial segment of reference speech is given to the decoder, and the remaining linguistic information is forced to be extracted from the phone sequence.

During inference, our system supports style reference audio $\vb x_{sty}$ and speaker reference audio $\vb x_{spk}$ from different speakers and with different content.
Moreover, we can adopt a similar sampling method in GST~\cite{Wang2018StyleTU} to sample style embeddings without the style reference speech $\vb x_{sty}$.
In particular, we sample uniformly from the LST codebook independently for each time step of the style embedding.
We then multiply the style embedding by a scale factor $\beta=0.25$.
The length of the style embedding is chosen randomly from $[80, 160]$.

\section{Experiment Setup}
We conduct our experiments under both single-speaker and multi-speaker settings, each designed to evaluate different aspects of our system.
We use two datasets, the LJ Speech dataset~\cite{ljspeech17} and the VCTK Corpus~\cite{VCTK} respectively.
The LJ Speech dataset~\cite{ljspeech17} is a commonly used dataset for single-speaker TTS systems.
It contains 13,100 English utterances from a single female speaker with total duration of about 24 hours.
The VCTK Corpus consists of 110 English speakers each reading about 400 sentences.
As various accents are included, it is often adopted for evaluation of voice conversion systems.
For VCTK, 44 male and female speakers were randomly selected for the training set, with the remaining speakers as test set.
Additionally, we use the ESD database~\cite{9413391}, an emotional speech synthesis corpus to evaluate style transferability in our system.
We only use the English corpus in ESD which consists of 13 hours of speech from 10 speakers with 5 emotions.

We extract mel-spectrograms from audio following the convention of Tacotron2~\cite{tacotron2}.
WaveGlow~\cite{waveglow} is used as the vocoder to transform mel-spectrogram to synthetic speech.
We use 64 GSTs and 32 LSTs and set the embedding and hidden size of all modules to 256 except the bottleneck layer of the decoder pre-net (32) and the FFN in attention blocks (1024).
We set the number of content-style cross attention blocks and transformer decoder blocks to $N=5$.
The system is trained using Adam optimizer~\cite{kingma2014adam} with learning rate $2\times 10^{-4}$.
We first train our system on LJ Speech for 250k steps with a batch size of 128.
The resulting model is used for single-speaker evaluation.
Since the speaker embedding is not useful for the single-speaker system, we simply use the average speaker embedding of the training examples at inference time to eliminate the need of speaker reference audio.
Bootstrapping from the pretrained model on LJ Speech, we further train our system for 150k steps on VCTK and ESD with a batch size of 64.
Additional implementation details are in our github repository.

\textbf{Comparing methods.}
We evaluate four different variants of our LST-TTS system.
SPK is a version which only uses speaker embedding without style embedding, this is similar to MultiSpeech~\cite{Chen2020MultiSpeechMT} in VCTK and TransformerTTS~\cite{Li2019NeuralSS} in LJ Speech.
LST-A refers to the situation where the speaker and style reference speech are from the same speaker, while in LST-S they are shuffled and may be from different speakers.
In LST-R, random sampling using the method described in Section~\ref{ssec:training} is applied for LST without conditioning on any reference speech.
Additionally, we denote ground truth speech in both dataset as GT.



\begin{table*}[th]
  \centering
  \caption{Objective evaluation on LJ Speech and VCTK}
  \label{tab:obj-exp}
  \centering
  \begin{tabular}{l c c c c c c c c c c}
    \toprule
    & \multicolumn{5}{c}{\textbf{LJ Speech}} & \multicolumn{5}{c}{\textbf{VCTK}}\\
    \cmidrule(lr){2-6} \cmidrule(lr){7-11}
    & GT & LST-A & Tan et al.~\cite{tan21_interspeech} & SPK & LST-R & GT & LST-A & UTTS~\cite{paul21_interspeech} & SPK & LST-S\\
    \midrule
    WER(\%) & 3.6 & 9.5 & 21.4 & 34.5 & 8.7 & 3.0 & 12.5 & 18.3 & 40.5 & 12.3\\
    \bottomrule
  \end{tabular}
\end{table*}
\vspace{-0.7em}
\section{Results and discussions}
\label{sec:result}
\subsection{Objective evaluation}
We first evaluate the systems using the word error rate (WER) obtained from an ASR model to measure content integrity.
We use the ASR model pretrained by ESPNet~\cite{watanabe2018espnet} on the LibriSpeech dataset~\cite{7178964}.
Table~\ref{tab:obj-exp} presents the WER of competing methods on both datasets.
In LJ Speech, our methods LST-A and LST-R both yield a low WER, which verifies that content-leakage is minimized in our system and ensures the intelligibility of the synthetic speech.
We also show that randomly sample the LST without sequential dependency (LST-R) does not affect the intelligibility of the generated speech.
This enables our system to sample different styles without the presence of $\vb x_{sty}$.
In addition, we observe that the SPK variant of our system tends to suffer from deletion and repetition, which causes the high WER.
We conjecture that with our style network and content-style cross-attention module, the content representation is adjusted to contain the prosody information.
As a result, our decoder network can focus on learning the alignment without modeling the speaking style, leading to better robustness.

We further show on VCTK the ability of our system to scale to multi-speaker settings.
Table~\ref{tab:obj-exp} shows that both LST-A and LST-S are able achieve low WER, and that passing different speakers for $\vb x_{sty}$ and $\vb x_{spk}$ does not affect intelligibility.
While not directly comparable due to the difference in ASR models, we include the WER of recent TTS systems with style modeling~\cite{tan21_interspeech,paul21_interspeech} on each dataset for reference.

\textbf{Emotional TTS.}
We assess the style transferability of our method objectively using a pretrained speech emotion recognition (SER) classifier with 95.8\% accuracy on the test set of ESD.
We label synthesized audio of our model with the emotion label of the style reference speech $\vb x_{sty}$ and calculate the accuracy of the SER classifier over 1000 randomly sampled synthetic speech.
Under the LST-A setup, we achieved 68.4\% accuracy, which demonstrates emotion patterns can be transferred from $\vb x_{sty}$ with our system.
While being lower in performance compared to recent Emotional TTS (ETTS) systems~\cite{liu21p_interspeech,ETTSXiong,Li2021ControllableET}, our system can realize ETTS without any supervision from the emotion labels.

\vspace{-0.5em}
\subsection{Subjective evaluation}
We evaluate the naturalness and style transferability of our system in terms of a listening test.
All tests were conducted using Amazon Mechanical Turk.
For naturalness, each competing method generated 20 utterances with text and reference speech randomly sampled from the test set.
Workers were asked to rate the naturalness of the speech on a 5-scale from 1 (Bad) to 5 (Excellent).
Each audio was rated 20 times by different workers and we report the mean opinion score (MOS) along with the 95\% confidence interval.
For style similarity, 20 pairs of style reference speech and synthesized speech was presented to the workers for each system.
Similarly, the workers rated the speaking style similarity between the two audios from 1 (not similar at all) to 5 (highly similar).

Table~\ref{tab:mos-lj} presents the evaluation result on LJ Speech.
With the TransformerTTS architecture, SPK already achieves a high naturalness score.
Adding the style control with our method does not degrade but improves the naturalness, suggesting the success of the fusion between linguistic content and speaking style by cross-attention module.
Compared to \cite{tan21_interspeech} which applied fine-grained style control on Tacotron2, our method outperforms in both naturalness and style similarity.
Table~\ref{tab:mos-vc} indicates the speech quality of our method is not impaired under multi-speaker configuration.
While LST-S lowers the similarity score, it could be attributed to the increase in difficulty of comparing speaking styles from different speakers.
On the other hand, for LST-A, the style similarity score in VCTK remains close to the that of LJ Speech.
These results demonstrate that our system is able to perform speaking style transfer within each speaker.

\begin{table}[th]
  \caption{MOS with 95\% confidence interval on LJ Speech}
  \label{tab:mos-lj}
  \centering
  \begin{tabular}{l c c c c}
    \toprule
    & Naturalness & Style similarity\\
    \midrule
    GT & $4.07\pm 0.07$ & - \\
    LST-A & $\textbf{3.62}\pm 0.09$ & $\textbf{3.28}\pm 0.10$ \\
    Tan et al.~\cite{tan21_interspeech} & $3.19\pm 0.11$ & $3.23\pm 0.11$ \\
    SPK & $3.50\pm 0.11$ & - \\
    \bottomrule
  \end{tabular}
\end{table}
\vspace{-1.2em}
\begin{table}[th]
  \caption{MOS with 95\% confidence interval on VCTK}
  \label{tab:mos-vc}
  \centering
  \begin{tabular}{l c c c c}
    \toprule
    & Naturalness & Style similarity\\
    \midrule
    GT & $3.87\pm 0.10$ & - \\
    LST-A & $\textbf{3.75}\pm 0.10$ & $\textbf{3.25}\pm 0.12$ \\
    LST-S & $3.73\pm 0.10$ & $3.09\pm 0.10$ \\
    SPK & $3.47\pm 0.10$ & $3.07\pm 0.13$\\
    \bottomrule
  \end{tabular}
\end{table}
\vspace{-1.5em}

\section{Conclusion}
\label{sec:conclusion}
We developed a transformer-based TTS system that incorporates fine-grained style control.
We evaluate the intelligibility, naturalness, and style transferability of the synthetic speech and show that our system can generalize to various settings.
We believe that the techniques we described are capable of abstracting key properties of speech and allowing these to be recombined to generate novel outputs.
Based on this work, we intend to inspect the learned fine-grained style embedding of our system and its potential downstream applications.

\vspace{-0.2em}
\section{Acknowledgements}
We thank Richard M. Stern and Raymond Xia for the discussion and feedback on our system.

\vfill\pagebreak

\let\OLDthebibliography\thebibliography
\renewcommand\thebibliography[1]{
  \OLDthebibliography{#1}
  \setlength{\parskip}{1.4pt}
  \setlength{\itemsep}{1.2pt}
}
\bibliographystyle{IEEEbib}
{\small\bibliography{main}}

\end{document}